\documentstyle[aaspp4]{article}

\def\be{\begin{equation}}
\def\ee{\end{equation}}
\def\ba{\begin{eqnarray}}
\def\ea{\end{eqnarray}}
\def\go{\mathrel{\raise.3ex\hbox{$>$}\mkern-14mu
             \lower0.6ex\hbox{$\sim$}}}
\def\lo{\mathrel{\raise.3ex\hbox{$<$}\mkern-14mu
             \lower0.6ex\hbox{$\sim$}}}
\def\cJ{{\cal J}}

\def\cO{{\cal O}}

\begin{document}

\title{Precession of Magnetically Driven Warped Disks and Low-Frequency
QPOs in Low-Mass X-Ray Binaries}
\author{Akiko Shirakawa and Dong Lai}
\affil{Center for Radiophysics and Space Research, Cornell University,
Ithaca, NY 14853\\
Email: shirak,dong@astro.cornell.edu}

\begin{abstract}
An accretion disk around a rotating magnetized star is subjected to
magnetic torques which induce disk warping and precession.
These torques arise generically from interactions between the stellar
field and the induced surface currents on the disk.
Applying these new effects to weakly magnetized ($B\sim$ $10^7$--$10^9$~G)
neutron stars in low-mass X-ray binaries, we study the global hydrodynamical
warping/precession modes of the disk under the combined influences of
relativistic frame dragging, classical precession due to the
oblateness of the neutron star, and the magnetic torques.
Under quite general conditions, the magnetic warping torque can overcome the
``Bardeen-Petterson'' viscous damping and makes the modes grow.
The modes are confined to the inner region of the disk, and have
frequencies equal to $0.3-0.95$ (depending on the mass accretion rate $\dot M$)
times the sum of the Lense-Thirring frequency, the classical precession
frequency, and the magnetically driven precession frequency evaluated at the
inner disk radius $r_{\rm in}$. As $\dot M$ increases, the mode frequency
is reduced relative to the total precession frequency at $r_{\rm in}$ since
the mode becomes less concentrated around $r_{\rm in}$ due to the increasing
viscous stress associated with the large $\dot M$. Because of this, and because
the magnetically driven precession is retrograde (opposite to
the Lense-Thirring precession) and depends strongly on $\dot M$,
the mode frequency can have a non-monotonic dependence on the mass
accretion rate. This may account for several observed features of low-frequency
(10--60~Hz) quasi-periodic oscillations (LFQPOs) in low-mass X-ray binaries
which are otherwise difficult to explain,
such as the flattening/turnover in the LFQPO frequency -- $\dot M$ correlation
or in the LFQPO frequency -- kHz QPO frequency correlation (e.g.,
as seen clearly in GX 17+2).

\end{abstract}


\section{Introduction}
Accreting neutron stars (NSs) in low-mass X-ray binaries (LMXBs) exhibit
rapid, quasi-periodic oscillations (QPOs) in their X-ray fluxes.
The so-called horizontal-branch oscillations (HBOs), discovered in a subclass
of LMXBs called Z sources,
are low-frequency QPOs (LFQPOs) with centroid frequencies in the range
of 15-60~Hz, $Q$-values of order a few, and rms amplitudes $\lo$ 10~\%
(see van der Klis 1995 for a review).
One interpretation of the HBOs is the
magnetospheric beat-frequency model (Alpar \& Shaham 1985;
Lamb et al.~1985) in which the HBO frequency is identified
with the difference frequency between the Keplerian frequency at the
magnetospheric boundary and the spin frequency of the NS.
However, over the last few years,
the discovery of kilohertz (300--1300~Hz) QPOs in $\sim 20$ LMXBs by the {\it
Rossi X-Ray Timing Explorer} ({\it RXTE}) has called into question this
interpretation of the HBOs (see van der Klis 2000 for a review). In particular,
observations indicate that in most Z sources, kHz QPOs
(which often come in pairs) occur simultaneously with the HBOs.
In several atoll sources (which are thought to have weaker magnetic fields
and smaller accretion rates than the Z sources), similar LFQPOs
have also been found at the same time when kHz QPOs occur.
While the origin of the kHz QPOs is still being debated, it is convenient
(as in all current theoretical models; see van der Klis 2000 and references
therein) to associate one of the kHz peaks
with the orbital motion in the inner edge of the accretion disk. On the other
hand, the discovery of nearly coherent oscillations during X-ray bursts in
several LMXBs establishes the spin frequency of the NS to be
$\sim$ 300--600~Hz (see Strohmayer 2000). Obviously, the beating between the
kHz orbital frequency and the spin frequency cannot produce the 10--60~Hz
LFQPOs.

Stella \& Vietri (1998) suggested that the LFQPOs are associated with
the Lense-Thirring precession of the inner accretion disk around the rotating
NS. If the LFQPO and the kHz QPOs are generated at the same special radius
in the disk, this
implies a quadratic relation between the LFQPO frequency and the kHz QPO
frequency, in rough agreement with observations
(usually for certain range of X-ray fluxes; but see \S 4.4).
For reasonable values of NS moment of inertia, the data require
that the observed LFQPO peaks correspond to 2 or 4 times the Lense-Thirring
frequency (Stella \& Vietri 1998; Ford \& van der Klis 1998;
Psaltis et al. 1999); this may be possible
since a precessing disk may induce X-ray flux variation mainly at the harmonics
of its fundamental precession frequency and there may be ``sub-harmonic''
feature buried in the X-ray power spectra (see Jonker et al.~2000).

There are several theoretical and observational puzzles
associated with the Lense-Thirring interpretation of LFQPOs.
First, it is well-known that the combined effects of differential precession
and viscosity tend to keep the inner region of the disk [within
$\sim (100-1000)GM/c^2$] perpendicular to the NS spin axis (Bardeen \&
Petterson 1975). A mechanism to drive the disk tilt is then needed for the
precession to make sense. Second, because different ``rings''
in the disk have different precession rates and are strongly coupled to each
other, a global mode analysis is needed to determine the true precession rate
of the inner disk. Markovi\'c \& Lamb (1998) carried out such an analysis but
found that all the modes are damped even when the radiation driven warping
torque (Pringle 1996) is included. Third, pure Lense-Thirring precession
has difficulty in explaining the observed behavior of
the LFQPO frequency as a function of the accretion rate of the system
(Wijnands et al.~1996; Psaltis et al.~1999; see \S 4.4).

In this paper, we show that the magnetically driven warping instability (Lai
1999), resulting from interactions between the weakly
magnetized ($\sim 10^8$~G) NS and the inner accretion disk (see \S 2),
can naturally overcome the Bardeen-Petterson viscous damping and therefore
induces a tilt in the inner disk. By carrying out a global analysis of disk
precession/warping modes, including magnetic torques
in addition to the Lense-Thirring and
classical precession torques, we obtain, for the first time,
growing warping/precession modes of the disk (see \S 3 and \S 4).
Moreover, we show that the magnetically driven precession torque
(Lai 1999; see \S 2) significantly
affects the actual disk precession frequency as the accretion rate increases;
this can potentially explain the observed $\dot M$-dependent behaviors
of the LFQPOs (\S4).

\section{Magnetically Driven Warping and Precession}

It is known that if the accretion disk is slightly tilted
from the equatorial plane of a rotating neutron star, it precesses
due to general relativistic frame-dragging effect (Lense-Thirring precession)
and also due to oblateness of the rotating star (classical precession).
For a magnetized neutron star, there is an additional magnetically driven
precession and warping of the disk; these arise from interactions between the
stellar magnetic field and the disk before the latter is truncated at the
magnetospheric boundary (Lai 1999). The magnetic torques appear
under the generic condition that the
stellar magnetic field (treated as a dipole in this paper)
is not aligned with the stellar spin axis.
Depending on how the disk responds to the stellar field,
two different kinds of torque arise:
(i) If the vertical stellar magnetic field $B_z$ penetrates the disk,
it gets twisted by the disk rotation
to produce an azimuthal field $\Delta B_\phi=\mp\zeta B_z$ that has different
signs above and below the disk ($\zeta$ is the azimuthal pitch of the field
line and depends on the dissipation in the disk), and a radial surface current
$K_r$ results. The interaction between $K_r$ and the stellar $B_\phi$ gives
rise to a vertical force. While the mean force (averaging over the azimuthal
direction) is zero, the uneven distribution of the force induces a net
{\it warping torque} which tends to misalign the angular momentum of the disk
with the stellar spin axis.
(ii) If the disk does not allow the vertical stellar field
(e.g., the rapidly varying component of $B_z$ due to stellar rotation)
to penetrate, an azimuthal screening current $K_\phi$ will be induced on the
disk. This $K_\phi$ interacts with the radial magnetic field $B_r$
and produces a vertical force. The resulting {\it precessional torque}
tends to drive the disk into retrograde precession around the stellar spin
axis.

In general, both the magnetic warping torque and the precessional torque are present.
For small disk tilt angle $\beta$ (the angle between the disk normal and the
spin axis), the warping rate and precession angular frequency
at radius $r$ are given by
(see Lai 1999)
\be
\Gamma_m (r)=\frac{\zeta\mu^2}{4\pi r^7\Omega(r)\Sigma(r)}\cos^2\theta,
\label{eqn:Gamma_m}
\ee
and
\be
\Omega_m (r)=-\frac{\mu^2}{\pi^2 r^7\Omega(r)\Sigma(r) D(r)}\sin^2\theta,
\label{eqn:Omega_m}
\ee
where $\mu$ is the stellar magnetic dipole moment, $\theta$ is
the angle between the magnetic dipole axis and the spin axis,
$\Omega(r)$ is the orbital angular frequency, and $\Sigma(r)$ is the surface
density of the disk. The dimensionless function $D(r)$ is given by
\be
D(r)={\rm max}~(\sqrt{r^2/r^2_{\rm in}-1}, \sqrt{2H(r)/r_{\rm in}}),\label{eqn:D}
\ee
where $H(r)$ is the half-thickness and $r_{\rm in}$ is the inner radius of the disk.

Here we apply these magnetic torques to LMXBs containing weakly magnetized
($B\sim 10^8$~G) NSs. We parametrize the properties of the inner disk using
the ``inner region'' (radiation- and-scattering dominated)
solution of $\alpha$-disk (Shakura \& Sunyaev 1973)\footnote{Such a disk
is unstable against thermal-viscous perturbations (see Frank et al.~1992),
although the situation is not clear when there are magnetic fields 
threading the disk. Note that our main results related to the 
dimensionless mode frequencies (see \S 4.1 and \S 4.2) 
are not very sensitive to disk models (see
Shirakawa \& Lai 2001, where general power-law disks are studied).}, with
\ba
\Sigma(r)&=&(1050~{\rm g}~c{\rm m}^{-2})\alpha_{-1}^{-1}M_{1.4}^{-1/2}
{\dot M}_{17}^{-1}r_6^{3/2}\cJ(r)^{-1},
\label{eqn:surface density}\\
H(r)&=&(1.1~{\rm km}){\dot M}_{17}\cJ(r),
\label{eqn:half-thickness}
\ea
where $\alpha=(0.1)\alpha_{-1}$ is the $\alpha$-viscosity, $M=(1.4M_{\odot})M_{1.4}$ is
neutron star's mass, ${\dot M}=(10^{17}\,{\rm g\,s^{-1}}){\dot M_{17}}$ is
the mass accretion rate and $r_6=r/(10^6\,{\rm cm})$.
The dimensionless function $\cJ(r)$ is given by
\footnote{Magnetic fields threading the disk can modify $\cJ(r)$ in a model-dependent
way (see Lai 1999 for an example). However, the basic feature can still be
approximated by eq. (\ref{eqn:J}).}
\be
\cJ(r)=1-\xi\sqrt{\frac{r_{\rm in}}{r}}\label{eqn:J},
\ee
where $\xi$ is a dimensionless parameter with $0<\xi<1$.
Substituting equations (\ref{eqn:surface density}) and (\ref{eqn:half-thickness}) into
equations (\ref{eqn:Gamma_m}) and (\ref{eqn:Omega_m}), and assuming the Keplerian
flow for $\Omega(r)$, we get
\ba
\Gamma_m(r)&=&(55.3~{\rm s}^{-1})\,\zeta\cos^2 \theta \alpha_{-1}\mu_{26}^2
{\dot M}_{17}\cJ(r)\,r_6^{-7},
\label{eqn:Gamma_m2}\\
\Omega_m (r)&=&(-70.4~{\rm s}^{-1})\sin^2\theta\alpha_{-1}\mu_{26}^2{\dot M}_{17}
D(r)^{-1}\cJ(r)\,r_6^{-7},
\label{eqn:Omega_m2}
\ea
where $\mu_{26}=\mu/(10^{26}\,{\rm G\,cm}^3)$.

The Lense-Thirring precession angular frequency is given by
\footnote{In the strong gravitational field and/or rapid stellar
rotation regime, the Lense-Thirring precession (angular) frequency is replaced
by the nordal precession frequency $\Omega_{\rm nod}\equiv
\Omega-\Omega_{\theta}$ with $\Omega$ the orbital frequency and
$\Omega_{\theta}$ the vertical frequency.
The fractional deviation of $\Omega_{\rm nod}$ from $\Omega_{LT}$ is of order
${\hat a}(M/r)^{1/2}$ (where $\hat a=I\Omega_s/M^2$)
and will be neglected in this paper.}
\be
\Omega_{LT}(r)=\frac{2GI_3\Omega_s}{c^2 r^3}
=(280~s^{-1})I_{45}\left(\frac{\nu_s}{300~{\rm Hz}}\right)r_6^{-3},
\label{eqn:Omega_{LT}a}
\ee
where $I_3=(10^{45}~{\rm g}\,{\rm cm}^2)I_{45}$ is the moment of inertia
around the spin axis and $\Omega_s=2\pi\nu_s$ is the spin angular frequency.
The classical precession angular frequency is given by
\be
\Omega_{\rm cl}(r)=-\frac{3G(I_3-I_1)}{2r^3}\frac{\cos\beta}{\sqrt{GMr}},
\label{eqn:Omega_{cl}a}
\ee
where $I_1$ is the moment of inertia around the axis perpendicular to the spin.
If we approximate the rotating NS as a compressible Maclaurin spheroid
(Lai et al. 1993), we have $I_3=(1/5)\kappa_nM(a_1^2+a_2^2)$ and
$I_1=(1/5)\kappa_nM(a_2^2+a_3^2)$, where $a_i$ is the semi-major axis of the
spheroid, and $\kappa_n$ is a dimensionless parameter which depends on the
polytropic index (e.g., $\kappa_n$=1, 0.81482, 0.65345 for $n$=0, 0.5, 1,
respectively). For slow rotation, with
$\hat\Omega_s\equiv\Omega_s/(GM/R^3)^{1/2}\ll1$ (where $R$ is the radius of a nonrotating
star of the same mass), the semi-major axis is given by $a_i=R(1+\epsilon_i)$,
with (Lai et al. 1994; Appendix A)
\ba
\epsilon_1=\epsilon_2&=&\frac{1}{4}\kappa_n\left(1-\frac{n}{5}\right)\left(\frac{5+n}{3-n}\right)
{\hat \Omega_s}^2,\\
\epsilon_3&=&-\frac{1}{2}\kappa_n\left(1-\frac{n}{5}\right)\left(\frac{5-3n}{3-n}\right)
{\hat \Omega_s}^2.
\ea
Equations (\ref{eqn:Omega_{LT}a}) and (\ref{eqn:Omega_{cl}a}) then become (with $\cos\beta\simeq 1$)
\ba
\Omega_{LT}(r)&=&(313~{\rm s}^{-1})\kappa_n
\left[1+\frac{1}{2}\kappa_n\left(1-\frac{n}{5}\right)\left(\frac{5+n}{3-n}\right)
\hat\Omega_s^2\right]
\left(\frac{\nu_s}{300~{\rm Hz}}\right)M_{1.4}R_6^2r_6^{-3},
\label{eqn:Omega_{LT}}\\
\Omega_{\rm cl}(r)&=&(-195~{\rm s}^{-1})\kappa_n^2\left(1-\frac{n}{5}\right)
\left(\frac{\nu_s}{300~{\rm Hz}}\right)^2M_{1.4}^{-1/2}R_6^5r_6^{-7/2},
\label{eqn:Omega_{cl}}
\ea
where $R_6=R/(10^6\,{\rm cm})$. Morsink \& Stella (1999) have given exact numerical
solutions of $\Omega_{LT}$ and $\Omega_{\rm cl}$ for a variety of nuclear equations
of state. But analytic expressions (\ref{eqn:Omega_{LT}}) and (\ref{eqn:Omega_{cl}})
provide the dominant contributions, and are adequate for the purpose of this paper.

\section{Global Precession/Warping Modes: Equations}

Since the precession rates $\Omega_{LT},\Omega_{\rm cl},\Omega_m$
depend strongly on $r$, coupling between different rings is needed
to produce a global coherent precession. Such coupling can be achieved
through either viscous stress or bending waves (e.g., Papaloizou \& Pringle
1983; Terquem 1998 and references therein). In the viscosity dominated regime,
the formalism of Papaloizou \& Pringle (1983) can be used (see also Pringle
1992; Ogilvie 2000; Ogilvie \& Dubus 2001; note that in the linear regime, 
the different formalisms of disk warping dynamics are equivalent). 
We specify a warped precessing disk by the
disk normal vector ${\hat{\bf l}}(r,t)$. In the Cartesian coordinate,
with the $z$-axis along the NS spin, we write
${\hat{\bf l}}=(\sin\beta\cos\gamma,\sin\beta\sin\gamma,\cos\beta)$,
with $\beta(r,t)$ the tilt angle and $\gamma(r,t)$ the twist angle.
For $\beta\ll 1$, the dynamical warp equation for ${\hat{\bf l}}$
(see Lai 1999) reduces to an equation for $W(r,t)\equiv
\beta(r,t)e^{i\gamma(r,t)}$:
\be
\frac{\partial W}{\partial t}-
\left[\frac{3\nu_2}{4r}\left(1+\frac{2r\cJ'}{3\cJ}\right)
+\frac{3\nu_1}{2r}(\cJ^{-1}-1)\right]
\frac{\partial W}{\partial r}
=\frac{1}{2}\nu_{2}\frac{\partial^2 W}{\partial r^2}
+i\left(\Omega_{LT}+\Omega_{\rm cl}+\Omega_{m}\right)W+\Gamma_{m}W,\label{eqn:evolution}
\ee
where $\nu_2$ is the viscosity which tends to reduce the disk tilt.
We assume that the ratio of $\nu_2$ to the usual viscosity $\nu_1$ is constant.
For $\alpha$ disk, $\nu_1=\alpha H^2\Omega$, the viscosity rate is
\be
\tau_{\rm visc}^{-1}(r)\equiv{\nu_2(r)\over r^2}
=(16.7~{\rm s}^{-1})\,\left(\frac{\nu_2}{\nu_1}\right)\,
\alpha_{-1}M_{1.4}^{1/2}{\dot M_{17}}^2\cJ(r)^2\,r_6^{-7/2}.
\label{eqn:visc_rate}
\ee
To look for global modes we consider solutions of the form
$W(r,t)=e^{i\sigma t}W(r)$ with the complex mode frequency $\sigma$ ($=\sigma_r+i\sigma_i$).
It is convenient to define a dimensionless mode frequency
$\hat\sigma=\sigma\tau_{\rm visc}(r_{\rm in})$, where $\tau_{\rm visc}(r_{\rm in})$
is given by eq.~(\ref{eqn:visc_rate}) evaluated at $r_{\rm in}$, the inner disk radius.
Similarly, we define dimensionless quantities:
\ba
\hat\Omega_{LT}&\equiv&\Omega_{LT}(r_{\rm in})\tau_{\rm visc}(r_{\rm in})
=16.8\,I_{45}\left(\frac{\nu_s}{300~{\rm Hz}}\right)
\left(\frac{\nu_1}{\nu_2}\right)\alpha_{-1}^{-1}M_{1.4}^{-1/2}{\dot M_{17}}^{-2}
\cJ_{\rm in}^{-2}\left(\frac{r_{\rm in}}{10 {\rm km}}\right)^{1/2},
\label{eqn:hatOmega_{LT}}\\
\hat\Omega_{\rm cl}&\equiv&\Omega_{\rm cl}(r_{\rm in})\tau_{\rm visc}(r_{\rm in})
=-11.7\,\kappa_n^2\left(1-\frac{n}{5}\right)\left(\frac{\nu_s}{300~{\rm Hz}}\right)^2
\left(\frac{\nu_1}{\nu_2}\right)\alpha_{-1}^{-1}M_{1.4}^{-1}{\dot M_{17}}^{-2}R_6^5
\cJ_{\rm in}^{-2},
\label{eqn:hatOmega_{cl}}\\
\hat\Omega_m&\equiv&\Omega_m(r_{\rm in})\tau_{\rm visc}(r_{\rm in})
=-4.2\,\sin^2\theta\left(\frac{\nu_1}{\nu_2}\right)
\mu_{26}^2M_{1.4}^{-1/2}{\dot M_{17}}^{-1}D_{\rm in}^{-1}\cJ_{\rm in}^{-1}
\left(\frac{r_{\rm in}}{10 {\rm km}}\right)^{-7/2},
\label{eqn:hatOmega_m}\\
\hat\Gamma_m&\equiv&\Gamma_m(r_{\rm in})\tau_{\rm visc}(r_{\rm in})
=3.3\,\zeta \cos^2\theta \left(\frac{\nu_1}{\nu_2}\right)
\mu_{26}^2M_{1.4}^{-1/2}{\dot M_{17}}^{-1}
\cJ_{\rm in}^{-1}\left(\frac{r_{\rm in}}{10 {\rm km}}\right)^{-7/2},
\label{eqn:hatGamma_m}
\ea
where $D_{\rm in}\equiv D(r_{\rm in})$ and $\cJ_{\rm in}\equiv\cJ(r_{\rm in})$.
Note that we can use eq.
(\ref{eqn:Omega_{LT}}) to obtain more explicit expression of $\hat\Omega_{LT}$.
Equation (\ref{eqn:evolution}) can now be reduced to the dimensionless form:
\ba
&&i{\hat\sigma}W
-\left[\frac{3}{4x^{5/2}}\left(1+\frac{2x\cJ'}{3\cJ}\right)
+\frac{3}{2x^{5/2}}\left(\frac{\nu_1}{\nu_2}\right)(\frac{1}{\cJ}-1)\right]
\frac{\cJ^2}{\cJ_{\rm in}^2}\frac{dW}{dx}\nonumber\\
&&=\frac{1}{2x^{3/2}}\frac{\cJ^2}{\cJ_{\rm in}^2}\frac{d^2W}{dx^2}
+i\left[\frac{\hat\Omega_{LT}}{x^3}
+\frac{\hat\Omega_{\rm cl}}{x^{7/2}}
+\frac{\hat\Omega_m}{x^7}\frac{D_{\rm in}}{D}\frac{\cJ}{\cJ_{\rm in}}\right]W
+\frac{\hat\Gamma_m}{x^7}\frac{\cJ}{\cJ_{\rm in}}W,\label{eqn:oureqn2}
\ea
where $x\equiv r/r_{\rm in}$ and $\cJ'=d\cJ/dx$.

It is clear from eq.~(\ref{eqn:oureqn2}) that $\hat\sigma$ depends
only on five dimensionless parameters $\hat\Omega_{LT}$, $\hat\Omega_{\rm cl}$,
$\hat\Omega_m$, $\hat\Gamma_m$, and $\nu_1/\nu_2$ as well as two dimensionless functions
$D(x)$ and $\cJ(x)$.
To obtain $\sigma$ in physical units we need to know $r_{\rm in}$.
We adopt the simple ansatz:
\be
r_{\rm in}={\max}(r_m, r_{\rm ISCO}),
\label{eqn:r_{in}}
\ee
where the magnetosphere radius $r_m$ is given by
\be
r_m=18\,\eta\,\mu_{26}^{4/7}M_{1.4}^{-1/7}{\dot M}_{17}^{-2/7}~{\rm km},
\label{eqn:r_m}
\ee
(with $\eta\sim 0.5$), and the inner-most stable circular orbit
\footnote{The correction to $r_{\rm ISCO}$ due to stellar rotation is
negligible since $\hat a\sim 0.1$ for NSs in LMXBs.}
$r_{\rm ISCO}$ is given by
\be
r_{\rm ISCO}=6GM/c^2=12.4\,M_{1.4}~{\rm km}.
\label{eqn:r_{ISCO}}
\ee
The critical mass accretion rate $\dot M_{17,\,{\rm c}}$ below which $r_{\rm in}=r_m$
and above which $r_{\rm in}=r_{\rm ISCO}$ is obtained by equating
eqs.~(\ref{eqn:r_m}) and (\ref{eqn:r_{ISCO}}) as:
\be
\dot M_{17,\,{\rm c}}=3.7\,\eta^{7/2}\mu_{26}^2 M_{1.4}^{-4}.
\label{eqn:crit}
\ee
A more elaborate prescription for the inner disk radius when both general
relativity and the magnetic field are important is discussed in Lai (1998).

To solve eq.~(\ref{eqn:oureqn2}) for the complex eigenfunction $W(x)$ and
eigenvalue ${\hat \sigma}$, six real boundary conditions are needed.
In our calculation, the disk extends from $x_{\rm in}=1$ to $x_{\rm out}=50$.
For large $x$ and large $|\hat \sigma|$,
equation (\ref{eqn:oureqn2}) can be solved analytically, giving
\be
W(x)\propto\exp\,\left[\frac{4\sqrt{2}}{7}\,(i{\hat\sigma})^{1/2}
\cJ_{\rm in}x^{7/4}\right],
\ee
where we should choose the sign of $(i{\hat \sigma})^{1/2}$ so that
$W(x)\rightarrow 0$ as $x\rightarrow\infty$.
This approximate analytical solution, evaluated at $x_{\rm out}$,
together with its derivative, gives four (real) outer boundary conditions.
The inner boundary condition generally takes the form
$W'(x_{\rm in})=aW(x_{\rm in})$, with $a$ being a constant.
Most of our results in \S 4 will be based on $a=0$ (corresponding
to zero torque at the inner edge of the disk), although we
have experimented with different $a$'s (see Figs.~1-2)
and found that for $|a|\lo 1$ our results are unchanged to the extent that
there are other parameters in the problem which have greater uncertainties.
In numerically searching a
mode, we make a guess for the eigenvalue ${\hat \sigma}$ and
integrate eq.~(\ref{eqn:oureqn2}) from $x_{\rm out}$ to $x_{\rm in}$ using
the Kaps-Rentrop scheme (Press et al.~1992). We use
the globally convergent Newton method to find the correct value of ${\hat
\sigma}$ that satisfies the boundary conditions.

\section{Numerical Results and Discussion}

In this section, we first study numerical properties of equation (\ref{eqn:oureqn2})
and then discuss specific cases relevant to accreting NSs in LMXBs.

\subsection{Mode Eigenfunction and Eigenvalue}
For a given set of parameters $(\hat\Omega_{LT},\hat\Omega_{\rm cl},\hat\Omega_m,
\hat\Gamma_m)$, equation (\ref{eqn:oureqn2}) allows for many eigenmodes.
Here we shall focus on the ``fundamental'' mode which is more concentrated
near the inner edge of the disk and has larger $\hat\sigma_r$
(global precession frequency) and smaller $\hat\sigma_i$ (damping rate)
than any other ``higher-order'' modes (see Fig.~1).

If a mode were infinitely concentrated at the inner radius of the disk,
one expects that the mode frequency $\hat\sigma_r$
is just the sum of the frequencies evaluated at the inner disk radius,
i.e., $\hat\sigma_r=\hat\Omega_{LT}+\hat\Omega_{\rm cl}+\hat\Omega_m$.
However, the calculated $\hat\sigma_r$ is always smaller than
$\hat\Omega_{LT}+\hat\Omega_{\rm cl}+\hat\Omega_m$
because the mode is not infinitely concentrated but has a finite width.

Figure 1 shows the tilt angle $\beta(x,t=0)=|W(x)|$ associated with the modes
for different sets of ($\hat\Omega_{LT}$,$\hat\Gamma_m$).
We have set $\hat\Omega_{\rm cl}=\hat\Omega_m=0$ (since they play a similar
role as $\hat\Omega_{LT}$), $\cJ(x)=1$, and $\nu_1/\nu_2=1$
in eq.~(\ref{eqn:oureqn2}) for simplicity\footnote{
Note the results shown in Figs.~1-2 are not sensitive to ${\cJ}$ since 
$\cJ_{\rm in}$ has been absorbed in the definitions of dimensionless
frequencies, eqs.~(\ref{eqn:hatOmega_{LT}})-(\ref{eqn:hatGamma_m}). 
See also \S 4.3.}.
We see that as $\hat\Omega_{LT}$ and $\hat\Gamma_m$ increase,
the fundamental modes (solid lines) become more concentrated
near the inner radius of the disk.
This behavior can be understood heuristically: for a given $\Omega_{LT}(r)$,
a larger $\hat\Omega_{LT}$ implies smaller viscosity, and thus the coupling
between different disk radii is reduced.

Figure 2 shows the mode frequency $\sigma$ in units of $\Omega_{LT}(r_{\rm in})$,
or $\sigma/\Omega_{LT}(r_{\rm in})=\hat\sigma/\hat\Omega_{LT}$, as a function
of $\hat\Omega_{LT}$ for different values of $\hat\Gamma_m$ [We again set
$\hat\Omega_{\rm cl}=\hat\Omega_m=0$, $\cJ(x)=1$, and
$\nu_1/\nu_2=1$].
We see that $\hat\sigma_r/\hat\Omega_{LT}$ always lies between
0.4 to 0.95 for the relevant ranges of $\hat\Omega_{LT}$ ($10$ to $10^4$).
The ratio $\hat\sigma_r/\hat\Omega_{LT}$ increases and approaches unity as
$\hat\Omega_{LT}$ and $\hat\Gamma_m$ increase. This is consistent with the
behavior of the mode eigenfunction (see Fig.~1) that a larger $\hat\Omega_{LT}$
or $\hat\Gamma_m$ makes the mode more concentrated near the inner disk edge.

\subsection{Global Warping Instability Criterion}
As discussed in \S 1, differential precession tends to be damped by the $\nu_2$ viscosity.
Figure~2(b) shows that $\hat\sigma_i/\hat\Omega_{LT}=\sigma_i/\Omega_{LT}(r_{\rm in})$
is always positive (which implies damping)
for $\hat\Gamma_m=0$ and lies between $0.25$ (for
$\hat\Omega_{LT}=10$) and $0.04$ (for $\hat\Omega_{LT}=10^4$), corresponding to
$Q$ value ($Q\approx \sigma_r/\sigma_i$) between $2$ and $20$ (the high-$Q$ regime was also explored
by Markovic \& Lamb 1999 where a different disk model was adopted).
We see that $\hat\sigma_i$ decreases as $\hat\Gamma_m$ increases, and
becomes negative (implying mode growth) when the ratio
$\hat\Gamma_m/\hat\Omega_{LT}$ is sufficiently large.
The numerical values for $\hat\sigma_i$ can be approximated by
\be
\hat\sigma_i=-a\hat\Gamma_m+b{\hat\Omega_{LT}}^{0.7},
\ee
with $a\sim (0.5-1.0)$ and $b\sim 0.5$.
For the mode to grow ($\hat\sigma_i<0$) we require
\be
\hat\Gamma_m\go\hat\Omega_{LT}^{0.7}~\Longleftrightarrow~{\rm Global~Warping~Instability}.
\ee
For $\dot M_{17}<\dot M_{17,\,{\rm c}}$ [see eq. (\ref{eqn:crit})], so that
$r_{\rm in}=r_m$, this condition becomes
\be
0.7\,\zeta\,\mu_{26}^{-0.2}\cos^2\theta\,
\alpha_{-1}^{0.7}I_{45}^{-0.7}M_{1.4}^{0.4}{\dot M_{17}^{1.5}}
\left(\frac{\nu_s}{300~{\rm Hz}}\right)^{-0.7}\!\!
\left(\frac{\nu_1}{\nu_2}\right)^{0.3}\!
\left({\eta\over 0.5}\right)^{-3.85}\go 1.
\ee
For $\dot M_{17}>\dot M_{17,\,{\rm c}}$, so that $r_{\rm in}=r_{\rm ISCO}$,
we require
\be
0.2\,\zeta\,\mu_{26}^2\cos^2\theta\,
I_{45}^{-0.7}M_{1.4}^{-4}{\dot M}_{17}^{0.4}
\left(\frac{\nu_s}{300~{\rm Hz}}\right)^{-0.7}\!\!
\left(\frac{\nu_1}{\nu_2}\right)^{0.3}
\go 1.
\ee
We see that for parameters that characterize accreting NSs in LMXBs
the mode growth condition can be satisfied (see \S 4.3 for specific examples),
although not always.
In general, high (but not unreasonable) $\zeta$ ($>$ a few) and
$\dot M_{17}$ are preferred to obtain growing modes.

\subsection{Dependence of Mode Frequency on $\dot M$}
In \S 4.1 and \S 4.2, we have seen how $\hat\Omega_{LT}$ and $\hat\Gamma_m$
affect the dimensionless mode frequency $\hat\sigma$
($=\hat\sigma_r+i\hat\sigma_i$) while setting
$\hat\Omega_{\rm cl}=\hat\Omega_m=0$.
We have found that $\hat\sigma_r$ is smaller than the naively expected value
$\hat\Omega_{LT}$ for small $\hat\Omega_{LT}$ and $\hat\Gamma_m$,
but becomes closer to $\hat\Omega_{LT}$ as $\hat\Omega_{LT}$ and $\hat\Gamma_m$
increase. We have also found that a sufficiently large $\hat\Gamma_m$ tends to
make the mode unstable. In this subsection, we consider some specific examples
to illustrate the dependence of the mode frequency $\sigma$ on $\dot M$ and
other parameters for NSs in LMXBs.

Figure~3 shows the global precession frequency
(i.e., $\sigma_r/2\pi$; the real part of the mode frequency in physical units)
\footnote{In Figs.~3-6, we multiply the mode frequency by 4 to facilitate
comparison with observations, since the QPOs appear to manifest as 
harmonics of the fundamental frequency (Stella \& Vietri 1998; Psaltis et
al.~1999; see also \S 5).} as a function of the mass accretion rate for a fixed
parameter set [Parameter Set A: $M_{1.4}=1$, $R_6=1$, $n=1$, $\nu_s=300\,{\rm
Hz}$,
$\alpha_{-1}=1$, $\zeta=5$, $\sin^2\theta=0.1$, $\nu_2/\nu_1=1$, $\mu_{26}=2$,
$\eta=0.5$, $\xi=0$, and $D(x)$ as given in eq.~(\ref{eqn:D})], for three different cases:
(i) only the Lense-Thirring precession is included,
(ii) both the Lense-Thirring and classical precessions are included,
and (iii) the magnetic precession and warping are included in addition to
the Lense-Thirring and classical precessions. It also shows the
fiducial precession frequency, $\nu_{\rm fid}$, which is the sum
of the considered frequencies evaluated at the inner radius of the disk,
i.e., $\nu_{\rm fid}\equiv
(\Omega_{LT}(r_{\rm in})+\Omega_{\rm cl}(r_{\rm in})+\Omega_m(r_{\rm in}))/2\pi$.

The dependence of $\nu_{\rm fid}$ on $\dot M$
is understood in the following manner. From eqs.~(\ref{eqn:Omega_{LT}}), (\ref{eqn:Omega_{cl}}),
and (\ref{eqn:Omega_m}), together with eqs.~(\ref{eqn:r_{in}})--(\ref{eqn:r_{ISCO}}),
we find $\Omega_{LT}(r_{\rm in})\propto{\dot M}^{6/7}$,
$\Omega_{\rm cl}(r_{\rm in})\propto{\dot M}$, and $\Omega_m(r_{\rm in})\propto{\dot M}^3$
when $r_{\rm in}=r_m$, and $\Omega_{LT}(r_{\rm in})\propto{\dot M}^{0}$,
$\Omega_{\rm cl}(r_{\rm in})\propto{\dot M}^{0}$, and $\Omega_m(r_{\rm in})\propto{\dot M}$
when $r_{\rm in}=r_{\rm ISCO}$.
Thus we obtain, for the above three different cases:
\ba
&&({\rm i})
~\nu_{\rm fid}\propto\cO(\dot M^{6/7})~(\dot M_{17}<\dot M_{17,\,{\rm c}}),
~~\nu_{\rm fid}\propto\cO(\dot M^{0})~(\dot M_{17}>\dot M_{17,\,{\rm c}}).\label{eqn:fid1}\\
&&({\rm ii})
~\nu_{\rm fid}\propto\cO(\dot M^{6/7})-\cO(\dot M)~(\dot M_{17}<\dot M_{17,\,{\rm c}}),
~~\nu_{\rm fid}\propto\cO(\dot M^{0})~(\dot M_{17}>\dot M_{17,\,{\rm c}}).\label{eqn:fid2}\\
&&({\rm iii})
~\nu_{\rm fid}\propto\cO(\dot M^{6/7})-\cO(\dot M)-\cO(\dot M^3)~(\dot M_{17}<\dot M_{17,\,{\rm c}}),
\label{eqn:fid3a}\\
&&~~~~~~\nu_{\rm fid}\propto\cO(\dot M^{0})-\cO(\dot M)~(\dot M_{17}>\dot M_{17,\,{\rm c}})
\label{eqn:fid3b}.
\ea
For Parameter Set A adopted in Fig. 3,
the transition from $r_{\rm in}=r_m$ to $r_{\rm in}=r_{\rm ISCO}$
occurs at $\dot M_{17,\,{\rm c}}=1.3$ [see eq.~(\ref{eqn:crit})].

We see from Fig.~3 that the behavior of the mode frequency $\sigma_r/2\pi$
as a function of $\dot M$ is different from that of $\nu_{\rm fid}$
for all three cases (i)-(iii). For case (i) and (ii), $\sigma_r/2\pi$
is as small as $\sim30\%$ of $\nu_{\rm fid}$
for high $\dot M$ ($\dot M_{17}\sim1.5-2.0$) and the dependence of
$\sigma_r/2\pi$ on $\dot M$ is very different from that of $\nu_{\rm fid}$.
This is understood from our discussion in \S4.1:
The viscosity rate increases (and $\hat\Omega_{LT}+\hat\Omega_{\rm cl}$ decreases)
as $\dot M$ increases [see eq.~(\ref{eqn:visc_rate})]; this tends to spread
the mode by coupling (through viscous stress) disk rings at different radii,
thereby reducing $\sigma_r/2\pi$ relative to $\nu_{\rm fid}$.
The increasing importance of viscosity for high mass accretion rate is also expected
for case (iii), where additional magnetic effects are included.
However, we see from Fig.~3 that $\sigma_r/2\pi$ becomes closer to $\nu_{\rm fid}$
when the magnetic warping torque ($\Gamma_m$) is included. This is understood
from our finding in \S 4.1 (see Figs.~1--2) that the presence of $\Gamma_m$
makes the mode more concentrated near the disk inner radius.

Figure~4 shows the mode frequency $\sigma_r/2\pi$
as a function of $\dot M$ for three sets of parameters, Set A, B, and C,
where ($\mu_{26}$, $\sin^2\theta$) = (2, 0.1), (2, 0.5) and (4, 0.1), respectively,
while the other parameters are fixed to the standard values:
$M_{1.4}=1$, $R_6=1$, $n=1$, $\nu_s=300\,{\rm Hz}$, $\alpha_{-1}=1$,
$\zeta=5$, $\nu_2/\nu_1=1$, $\eta=0.5$, $\xi=0$.
All of the Lense-Thirring, classical precessions and magnetic precession and warping
are included [case (iii)].
For Parameter Set A and B, the inner radius of the disk
reaches $r_{\rm ISCO}$ at the same $\dot M_{17,\,{\rm c}}=1.3$
[see eq.~(\ref{eqn:crit})].
For Parameter Set A, $|\Omega_m(r_{\rm in})|$ is much smaller than
$\Omega_{LT}(r_{\rm in})+\Omega_{\rm cl}(r_{\rm in})$
for small $\dot M_{17}$ although it becomes more important for large $\dot M_{17}$;
thus at $\dot M_{17}=\dot M_{17,\,{\rm c}}$, $\sigma_r/2\pi$ suddenly changes
from an increasing function of $\dot M$ as given by eq.~(\ref{eqn:fid3a})
(with small second and third terms) to a linearly decreasing function
of $\dot M$ as given by eq.~(\ref{eqn:fid3b}).
For Parameter Set B, $|\Omega_m(r_{\rm in})| \propto \sin^2\theta$ is
5 times larger than that in Parameter Set A and is important
even for small $\dot M_{17}<\dot M_{17,\,{\rm c}}$;
thus $\sigma_r/2\pi$ shows a gradual turnover feature
according to eq.~(\ref{eqn:fid3a})
and becomes a linearly decreasing function
of $\dot M$ [see eq.~(\ref{eqn:fid3b})] above $\dot M_{17,\,{\rm c}}$.
For Parameter Set C, the inner radius of the disk is
always given by $r_m$ for the relevant $\dot M$ (since $\dot M_{17,\,{\rm c}}=5.2$).
As $|\Omega_{\rm cl}(r_{\rm in})|$ and $|\Omega_m(r_{\rm in})|$ are much smaller than
$\Omega_{LT}(r_{\rm in})$, $\sigma_r/2\pi$ is a monotonically
increasing function as given by eq.~(\ref{eqn:fid3a}) (with small second and
third terms).

Figure~5 illustrates the dependence of the mode frequency $\sigma_r/2\pi$ on the
polytropic index $n$ and the dimensionless parameter $\xi$
[defined in eq.~(\ref{eqn:J})] (other parameters are the same as in Parameter Set A
adopted in Fig.~4). Most nuclear equations of state have effective polytropic
index between 0.5 and 1. We see that $\sigma_r/2\pi$ is larger for a stiffer equation
of state (smaller $n$) and a larger $\xi$, although the dependences are not as strong
as on other parameters (cf, Fig.~4).

\subsection{Comparison with the Observed Behaviors of LFQPOs}

The behavior of $\sigma_r$ as a function of $\dot M$ (see Figs.~3--5)
is similar to the features observed for the LFQPOs in LMXBs (see van der Klis
1995, 2000).
While for some systems, the LFQPO frequencies increase as the X-ray flux
increases, for other systems, a non-monotonic (even opposite)
correlation between the QPO frequency and $\dot M$ has been observed.
As an example, for the Z source GX 17+2, using the position of the source on the X-ray
color-color diagram as an indicator of $\dot M$, Wijnands et al.~(1996)
found that the HBO frequency increases with $\dot M$ for small $\dot M$ but
decreases with further increase of $\dot M$.
In the simple model where the LFQPO is identified with the precession
frequency at the disk inner radius $r_{\rm in}$ (Stella \& Vietri 1998; see \S 1),
the non-monotonic behavior is difficult to explain, even when the classical
(retrograde) precession is included
\footnote{This is because $\Omega_{LT}\propto r^{-3}$ and
$\Omega_{\rm cl}\propto -r^{-7/2}$ have a similar $r$-dependence; see
eqs.~(\ref{eqn:Omega_{LT}}) and (\ref{eqn:Omega_{cl}}).}
(see Psaltis et al. 1999; Morsink \& Stella 1999).
However, we see from Fig.~4 that when we consider global disk modes and include
the magnetic effects, a variety of $\sigma_r-\dot M$ behaviors are possible.
In our model the turnover of the $\sigma_r-\dot M$ correlation occurs for two
reasons: (1) the magnetically driven retrograde precession 
$\Omega_m\propto -1/r^7$ becomes more important with increasing $\dot M$ [see
eqs.~(\ref{eqn:fid3a}) \& (\ref{eqn:fid3b})];
(2) the viscous stress becomes more important as $\dot M$ increases, and makes
the mode less concentrated around $r_{\rm in}$; thus $\sigma_r$ is
more reduced from $\nu_{\rm fid}$ with increasing $\dot M$. 
We note that even when magnetic effects are not included [case
(i) \& (ii)], the ``turnover'' feature is obtained due to mechanism (2) as
explained above (see Fig.~3). However, these modes are highly damped by
viscosity. Only the magnetic warping torque can make the mode grow (see
Fig.~4).

A related puzzle concerns the correlation
between the LFQPO frequency and the higher kilohertz QPO frequency
(e.g., Stella \& Vietri 1998; Ford \& van der Klis 1998; Psaltis et al.~1999;
see \S 1). It was found that for several sources
the positive correlation flattens (and even turns over) for large
kHz QPO frequencies (and large $\dot M$) (see Homan et al.~2001
for a clear case of such turnover in GX 17+2).
Again, this feature is difficult to explain with the simple picture of
Lense-Thirring and classical precessions at $r_{\rm in}$, but can be accounted
for qualitatively by our global disk oscillation model with magnetic effects
(see Fig.~6; note that we have not tried to vary the model parameters
to ``fit'' the observations). More quantitative comparison between our models
and the data require a better understanding of the origin of kHz QPOs (e.g., it
is not certain whether the kHz QPO frequency is the Keplerian frequency at
$r_{\rm in}$; see \S 5 for other uncertainties expected in real systems).

\section{Concluding Remarks}

We have shown in this paper that the inner region of the disk
around a weakly magnetized ($\sim 10^8$~G) NS is warped
due to magnetic field -- disk interactions
(see Lai 1999) and therefore must precess under the combined effects of
relativistic frame dragging, classical precession (due to the
oblateness of the NS) and magnetic torques. We have found
growing warping/precession modes of the inner disk and shown that
these modes have properties that resemble the 10--60~Hz low-frequency
QPOs observed in LMXBs (see \S 4).

Although our treatments of disk warping/precession go beyond
many previous models of QPOs (these models typically identify a QPO frequency
as the characteristic frequency of a test mass at certain special radius
in the disk; see van der Klis 2000 for a review), they still contain
important idealizations. For example, we have assumed the stellar field
to be dipolar when considering the magnetic field -- disk interactions.
This is unlikely to be correct for LMXBs since the disk
(with $r_{\rm in}$ only a few stellar radii) can produce significant
change in the global field topology even when the intrinsic stellar field is
dipolar (see Lai, Lovelace \& Wasserman 1999). Observationally, the
absence of persistent millisecond pulsation in all known LMXBs but
SAX J1808.4--3658 (Wijnands \& van der Klis 1998) and the frequency drift
of burst oscillations (e.g., Strohmayer 2000; see also Cumming et al.~2001)
may imply that the NS does not possess a well-defined dipole field,
although a more complex field is possible. Also, we have only studied the
linear behavior of the warping/precession modes in this paper.
Because of these idealizations and
the uncertainties in observations (e.g., it is difficult to determine
$\dot M$ precisely from observations), more quantitative comparison
between the observational data and our theory is premature at present.
Nevertheless, the results presented in this paper demonstrate that
magnetically driven warping and precession can give rise to a variety of
new possibilities for the dynamical behaviors of inner accretion disks
around magnetic NSs.

Concerning the observed properties of QPOs, many issues remain unanswered
in our paper (and in other related QPO studies): e.g, How does the
mode manifest itself as a variation in the X-ray flux?
How is the observed QPO amplitude (as large as $15\%$) produced?
One possibility is that the LFQPO is caused by occultation of
the central NS by the warped, precessing inner disk. Another issue is
that for the observed NS spin rates (based on observations of
burst oscillations) and reasonable moment of inertia of the neutron star,
the data require that the observed
LFQPO corresponds to 2 or 4 times the precession mode frequency (see \S1; the
fact that the global mode frequency is smaller than the precession frequency at
$r_{\rm in}$ exacerbates the problem); while this may be possible,
more theoretical study on this issue is needed. Finally, there is an indication
(but by no means proven) that some black hole systems display
low-frequency QPOs which resemble their counterparts in NS systems
(Psaltis, Belloni \& van der Klis 1999). Clearly, the magnetic
warping and precession torques discussed in this paper are irrelevant
to black hole systems\footnote{Note that even without
the magnetic torques and the classical precession, the mode frequency is
reduced relative to the Lense-Thirring frequency at $r_{\rm in}$ as
$\dot M$ increases (see Fig.~3); this result applies equally to the black hole
systems except that the mode is damped in the absence of other excitation
mechanisms.}. We note that the phenomenology
of black hole QPOs is much less developed, and that
many of the $\dot M$-dependent behaviors of LFQPOs observed in
NS systems are absent in the black hole systems (see Remillard 2001).

\acknowledgments
This work is supported in part by NSF Grant AST 9986740
and NASA grant NAG 5-8484, as well as by a research fellowship (to D.L.)
from the Alfred P. Sloan foundation.


\bigskip
\bigskip
\bigskip
\clearpage
\begin{figure}
\plotone{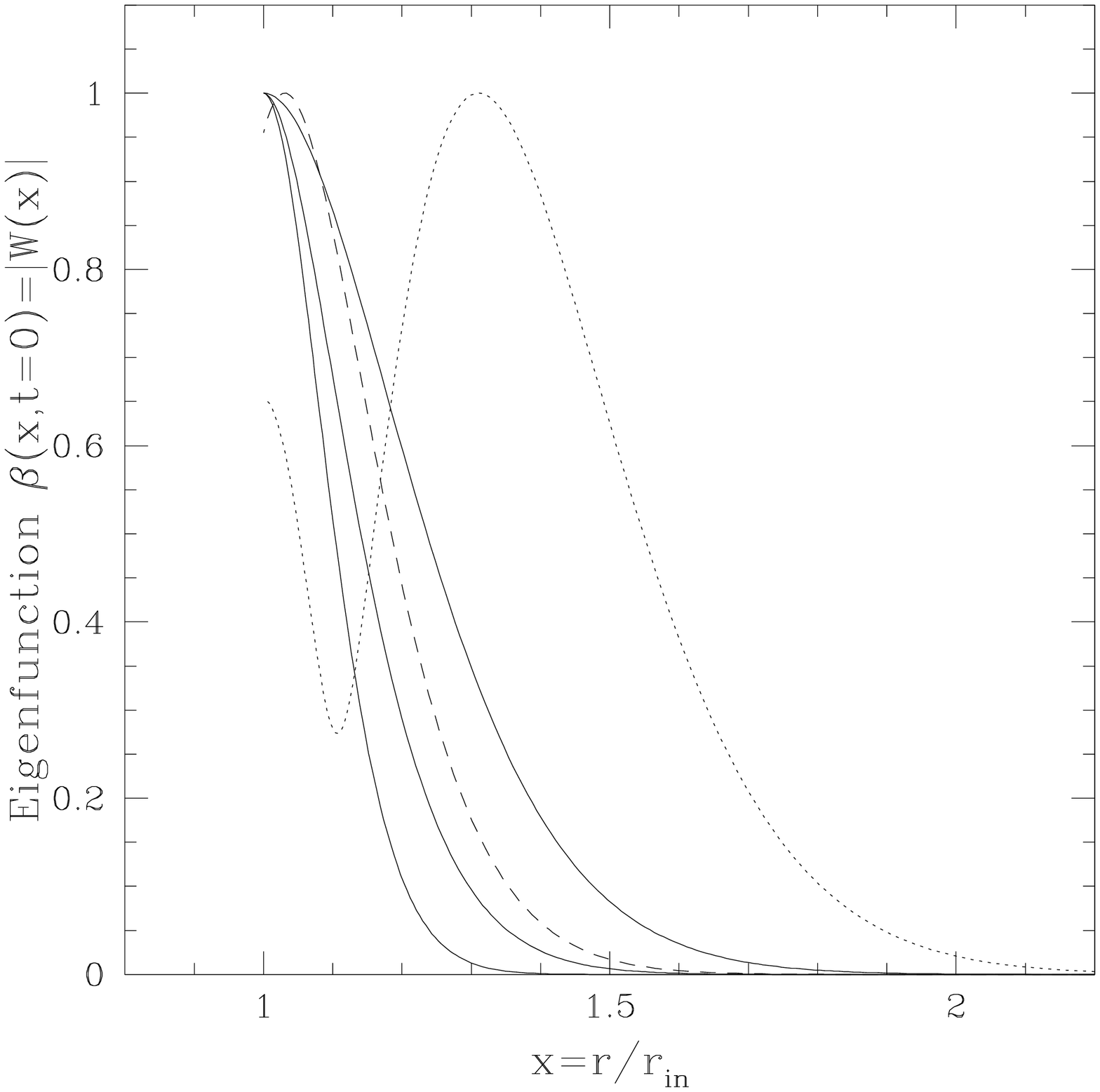}
\caption{The disk tilt angle of the warping/precession modes for
various parameter sets. The solid curves represent the fundamental modes for
($\hat\Omega_{LT}$, $\hat \Gamma_m$)
=(1000, 0), (100, 100), and (100, 0) (with $\hat\Omega_{\rm
cl}=\hat\Omega_m=0$) from left to right, with the corresponding mode frequency
$\hat\sigma=(\hat\sigma_r,\hat\sigma_i)=(860, 78)$, $(80, -46)$, and $(70,
15)$; the dotted curve represents a higher order mode
for ($\hat \Omega_{LT}$, $\hat\Gamma_m$)=(100, 100),
with $\hat\sigma=(\hat\sigma_r, \hat\sigma_i)=(41, 6.0)$; all these are
calculated using the inner boundary condition (B.C.) $W'=0$. The dashed curve
shows the case for ($\hat\Omega_{LT}$, $\hat \Gamma_m$)
=(100, 100), with the mode frequency $\hat\sigma=(65, -37)$, calculated
using the inner B.C. $W'=W$.
The eigenfunctions are normalized such that the maximum tilt angle is 1.
}
\end{figure}

\clearpage
\begin{figure}
\plotone{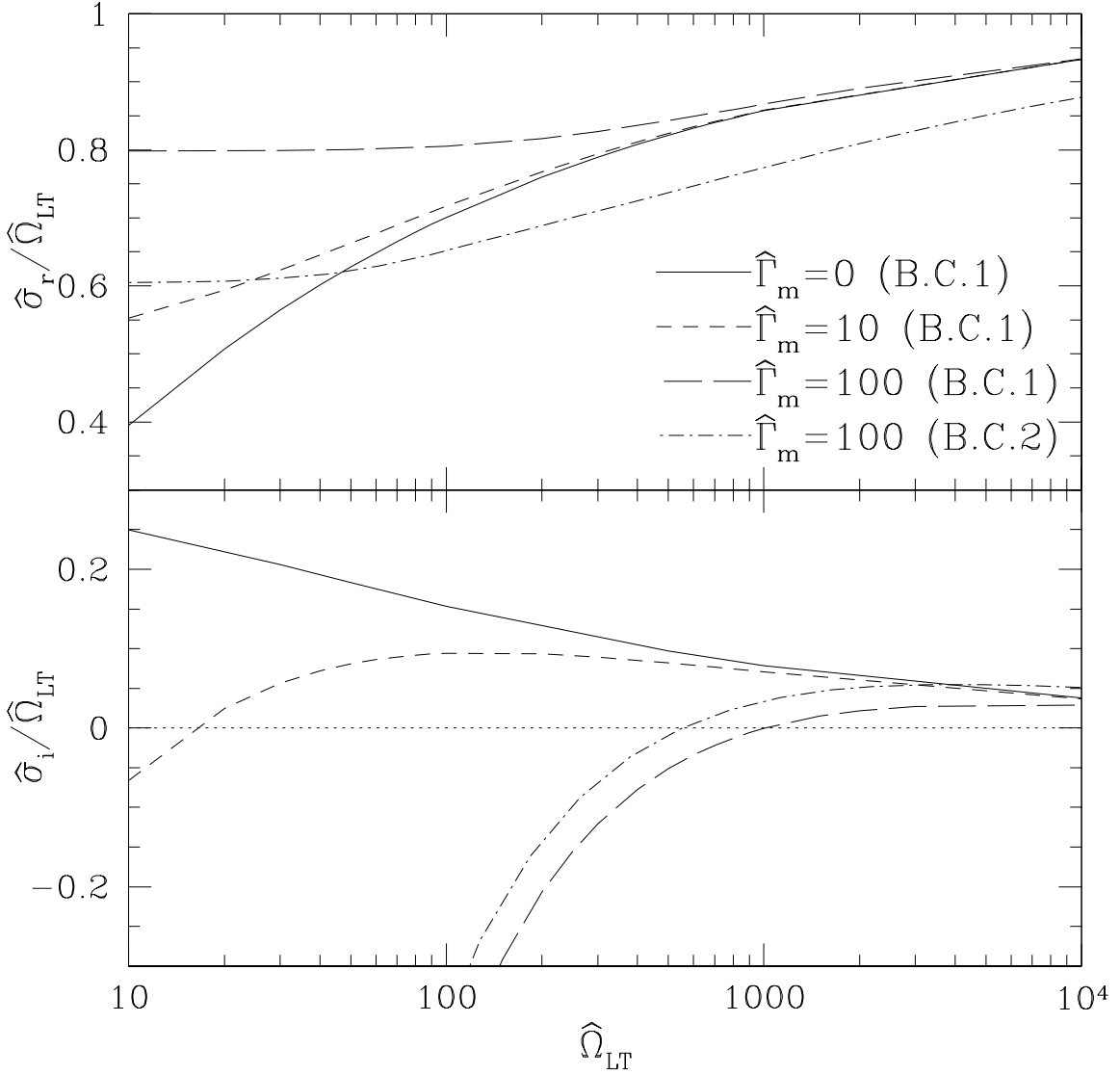}
\caption{
The upper panel shows the mode frequency $\hat\sigma_r$ in units of
$\hat\Omega_{LT}$ as a function of $\hat \Omega_{LT}$ for different values of
$\hat\Gamma_m$ (with $\hat\Omega_{\rm cl}=\hat\Omega_m=0$).
The lower panel shows the corresponding mode damping rate
$\hat\sigma_i$ (in units of $\hat \Omega_{LT}$).
Note that negative $\hat\sigma_i$ implies growing mode.
The dot-dashed curves are obtained using the inner
boundary condition $W'=W$, while all other curves are based on B.C.
$W'=0$.}
\end{figure}

\clearpage
\begin{figure}
\plotone{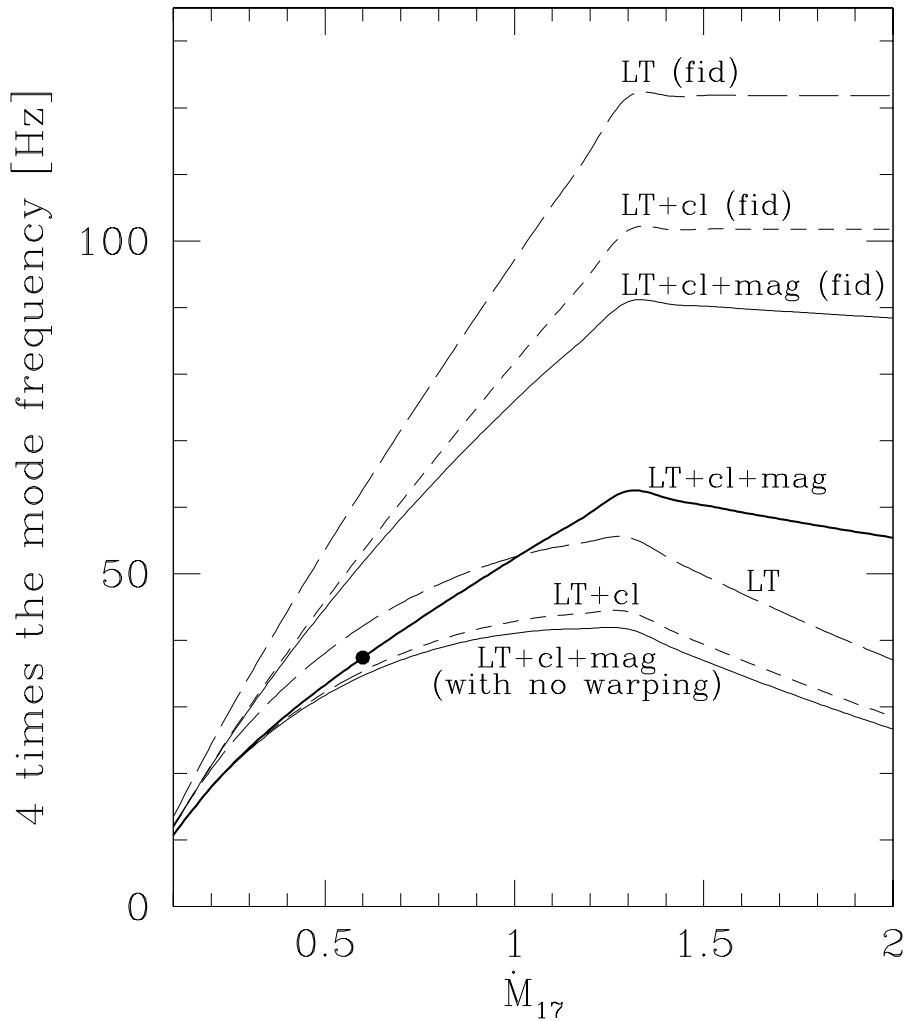}
\caption{The global precession mode frequency $\sigma_r/2\pi$ as a function of
the accretion rate ${\dot M_{17}}$. The frequencies are multiplied by 4 to
facilitate comparison with observations (see the text for discussion).
The upper three curves represent the
fiducial frequency $\nu_{\rm fid}$ for three different cases described in the text:
(i) only the Lense-Thirring precession is included (long dashed curve),
(ii) both the Lense-Thirring and classical precessions are included
(short dashed curve), and
(iii) the magnetic precession and warping are included as well as
the Lense-Thirring and classical precessions (solid curve). The thick solid curve is
the corresponding global precession frequency $\sigma_r/2\pi$ for case (iii)
(Note that the modes grow above $\dot M_{17}\simeq 0.6$; marked by a dot in the
figure).
The lower two dashed curves are the corresponding global precession frequency
$\sigma_r/2\pi$ for case (i) and (ii). The lower light solid curve shows
$\sigma_r/2\pi$ for case (iii) obtained with the magnetic warping torque
artificially turned off.
}
\end{figure}

\clearpage
\begin{figure}
\plotone{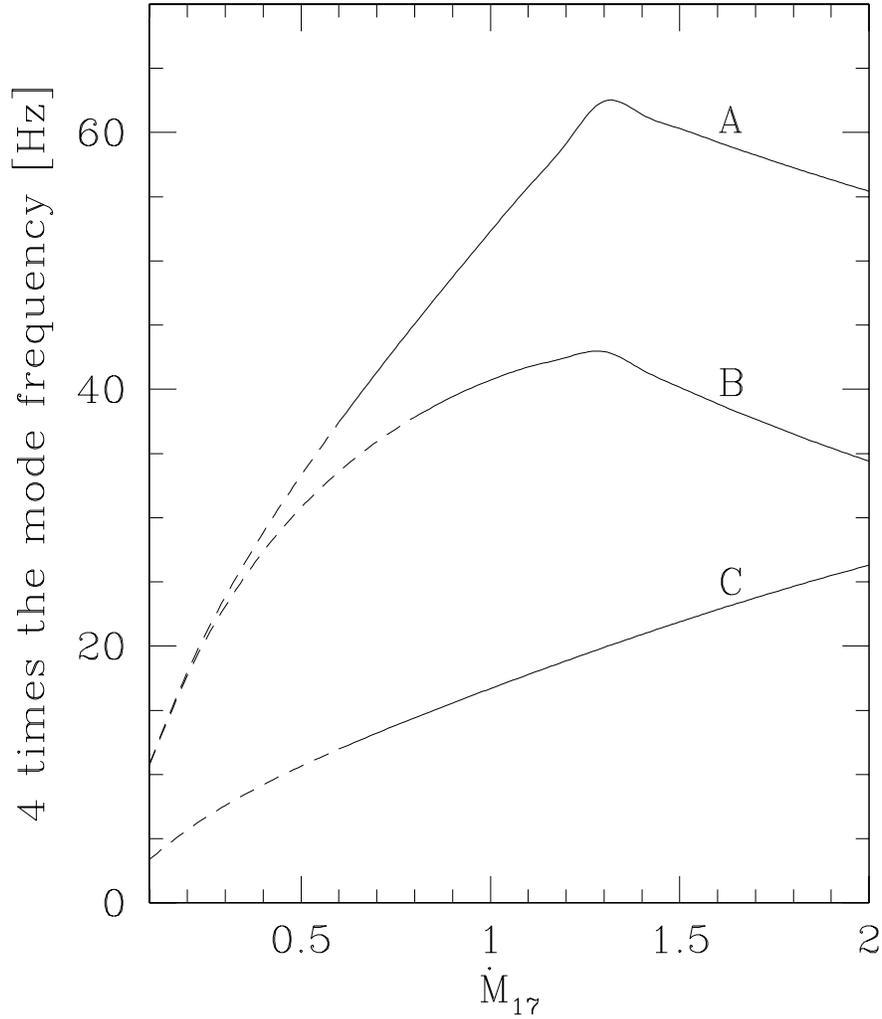}
\caption{Correlation between the mode frequency $\sigma_r/2\pi$
and the accretion rate ${\dot M_{17}}$ for Parameter Sets A, B, and C:
($\mu_{26}$, $\sin^2\theta$) = (2, 0.1), (2, 0.5), \& (4, 0.1), respectively
(see the text for the values of other fixed parameters)
when all of the Lense Thirring, classical, and magnetic precessions
and magnetic warping are taken into account.
The solid portion of the curve corresponds to growing mode and
the dashed portion corresponds to damping mode. To compare with LFQPOs
observed in LMXBs, we multiply $\sigma_r/2\pi$ by $4$.
}
\end{figure}

\clearpage
\begin{figure}
\plotone{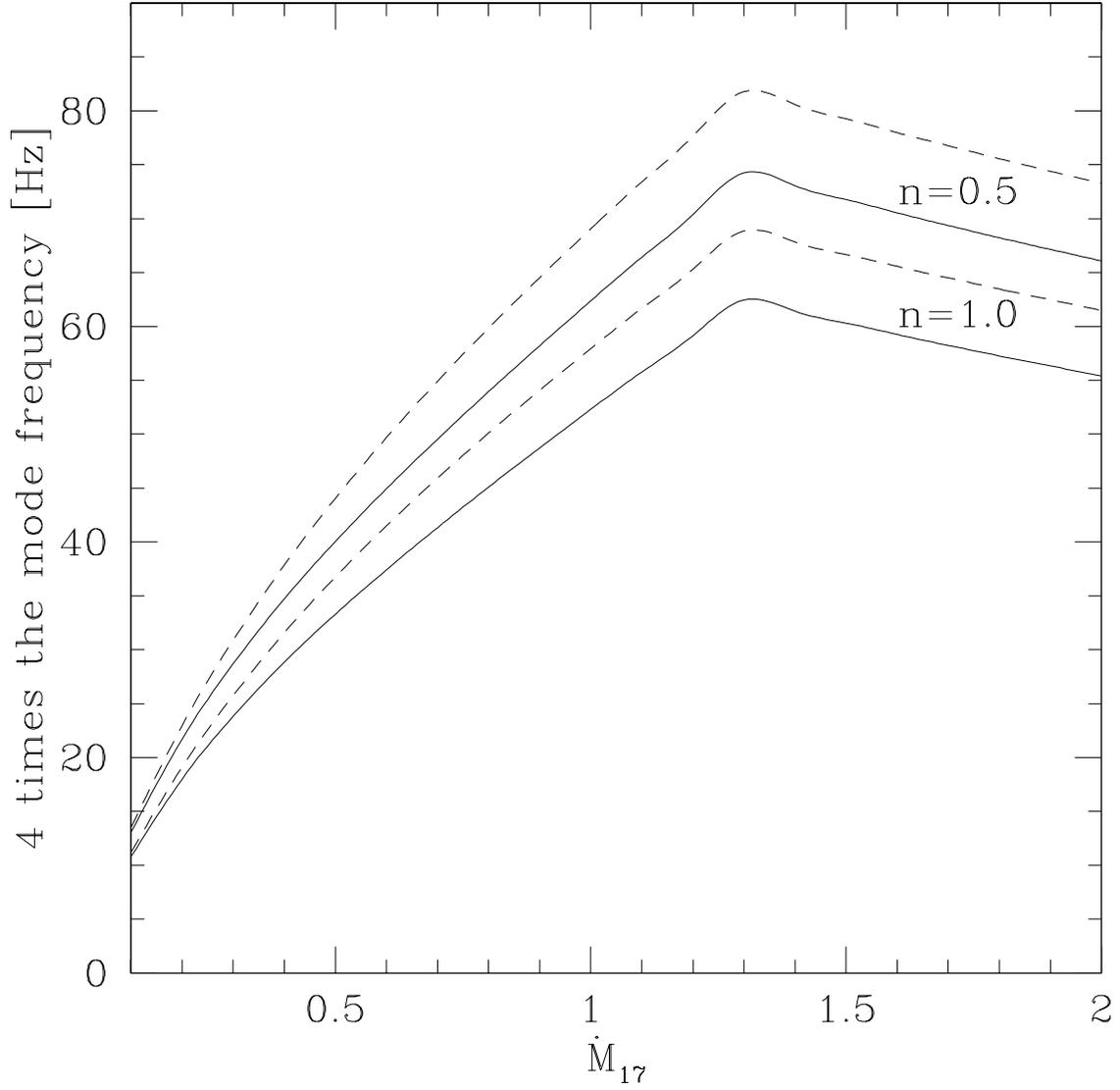}
\caption{The mode frequency $\sigma_r/2\pi$ (multiplied by $4$)
as a function of $\dot M_{17}$ for
polytrope index $n=0.5$ (upper two curves) and 1 (lower two curves), and
for $\xi=0$ (solid curves) and $\xi=0.5$ (dashed curves) [see
eq.~(\ref{eqn:J})]. Other parameters are the same as in Parameter Set A adopted
in Fig.~4.}
\end{figure}

\clearpage
\begin{figure}
\plotone{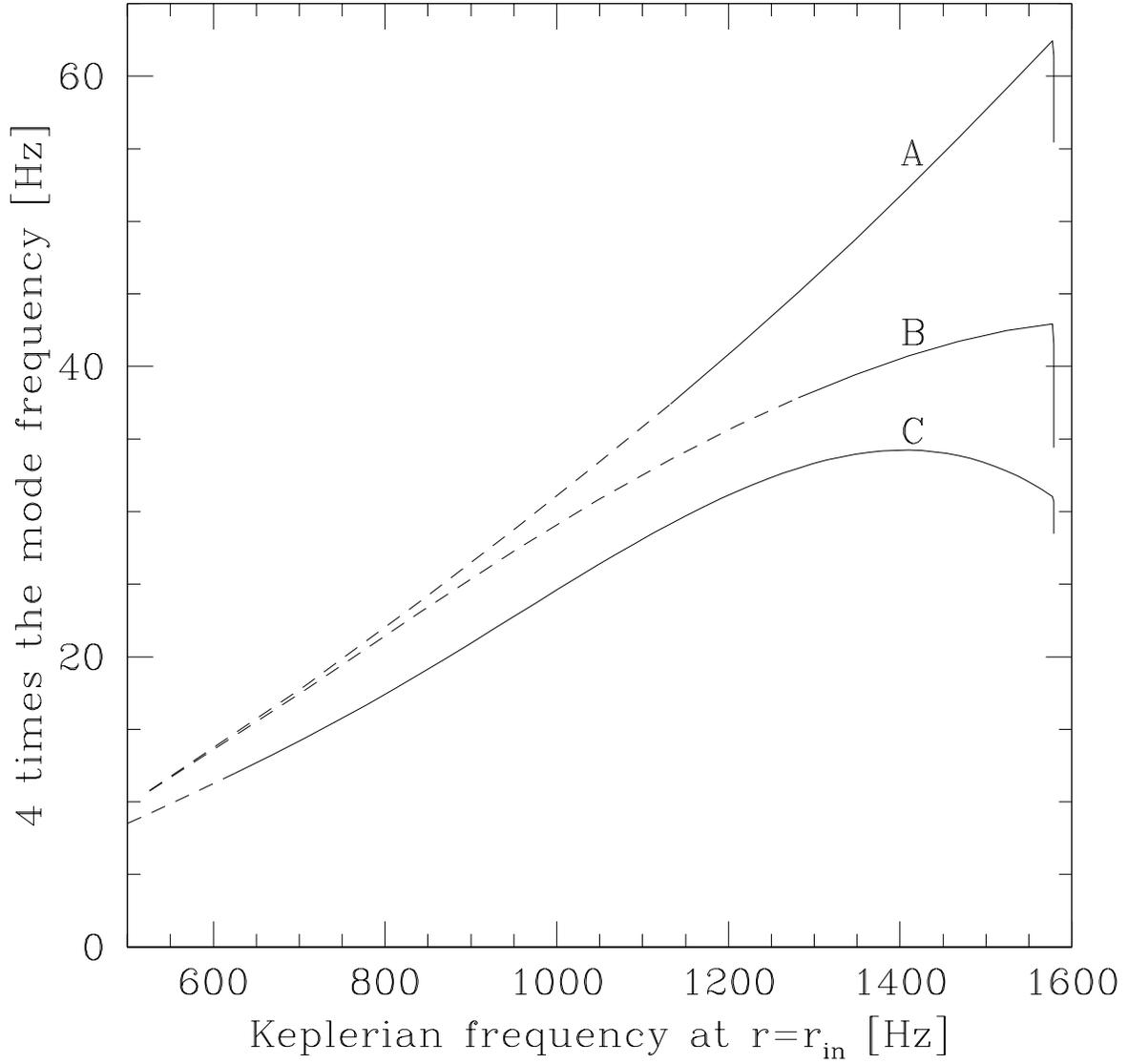}
\caption{Correlation between the mode frequency $\sigma_r/2\pi$ (multiplied
by $4$) and the Keplerian frequency at the inner radius of the disk for
Parameter Sets
A, B, and C as in Fig.~4. The solid portion of the curve corresponds
to growing mode and the dashed portion corresponds to damping mode.}
\end{figure}

\end{document}